# Conceptual Study of a Collective Thomson Scattering Diagnostic for SPARC

**Mads Mentz-Jørgensen, Riccardo Ragona[†], Søren B. Korsholm, Jesper Rasmussen**

Technical University of Denmark, Dept. of Physics, 2800 Lyngby, Denmark

Corresponding author: [†]`ricrag@fysik.dtu.dk`

**Abstract.** The SPARC tokamak is a compact high-field device that will operate at high plasma density with the aim to demonstrate net fusion energy. The experimentally unexplored plasma conditions in SPARC will require a carefully selected set of diagnostics for plasma monitoring and control. Here we explore conceptual design options and potential measurement capabilities of a collective Thomson scattering diagnostic at SPARC. We show that a 140 GHz X-mode CTS system is the most attractive option in terms of optimizing the signal-to-noise ratio and limiting sensitivity to refraction, as well as from a technological readiness perspective. Such a setup can provide core-localized measurements of the fusion alpha distribution function, main-ion temperature and toroidal rotation, fuel-ion ratio, and $^3$He content with relevant spatio-temporal resolution. Our proposed diagnostic layout can in principle be integrated into SPARC and could provide a valuable addition to its diagnostic suite at limited development costs and time.

## 1. Introduction

The demonstration of net fusion energy gain in magnetically confined plasmas is a fundamental step towards establishing the technical and commercial viability of fusion power plants. In recent years, multiple privately funded ventures have developed different concepts aimed at establishing a lead in the pursuit of integrating fusion energy into the global energy mix. Among these concepts, compact tokamaks with strong magnetic fields can potentially act as an enabling technology for achieving an economically attractive path towards a fusion power plant. In particular, the development of high-temperature superconductor technology paved the way for compact and affordable burning-plasma experiments with a goal of understanding reactor-relevant plasma regimes with net energy gain.

An example is the tokamak SPARC [1], which Commonwealth Fusion Systems aims to use as a stepping stone to commercial fusion energy [2]. In contrast to, e.g., ITER, SPARC is designed as a compact tokamak with small minor and major radii of $a = 0.57$ m and $R = 1.85$ m, respectively. Forgoing a large plasma volume, SPARC instead relies on operating with a strong on-axis magnetic field $B_0 \approx 12.5$ T using high-temperature superconducting magnets. A major aim is to achieve a net fusion energy gain $Q = P_{\text{fus}}/P_{\text{aux}} > 2$ [1]. This is partly facilitated by the small minor radius and the significant plasma current of $I_{\text{p}} \approx 8.7$ MA, which allow for a high Greenwald density limit and hence operation at core densities exceeding those in ITER and DEMO by a factor of $\sim 4$.

Such high-collisionality, high-$B$, $Q > 1$ plasma conditions represent an experimentally unexplored plasma regime, calling for diagnostics to characterize the plasma behaviour, monitor key parameters for plasma control, and preparing the technological and physics basis for a compact, high-field fusion power plant [3]. However, the implementation of diagnostics in a compact $Q > 1$ device presents unique challenges, including demands on the dimensions of the diagnostic infrastructure and its ease of integration due to the limited available space, as well as on its ability to operate in a harsh radiation environment. For these reasons, relevant diagnostics must be versatile, radiation resilient, and able to provide key measurements for physics and/or plasma control.

In this paper, we explore the feasibility and capability of a collective Thomson scattering (CTS) diagnostic in SPARC that meets these requirements. CTS offers a non-invasive tool to probe plasma properties at high spatial and temporal resolution, also in the plasma core and in reactor-relevant high-density plasmas [4, 5]. The technique has been employed at a range of devices, including JET [6], TEXTOR [7], FTU [8], ASDEX Upgrade [9], LHD [10], and Wendelstein 7-X [11]. A CTS diagnostic will be implemented in ITER [12] and is also being designed for DEMO [13].

CTS permits measurements of bulk- and fast-ion quantities such as bulk-ion temperature $T_{\text{i}}$, toroidal plasma rotation $v_{\text{tor}}$, distribution functions of fast ions such as fusion-born $\alpha$ particles, and plasma composition, including the fuel-ion ratio $R_{\text{i}} = n_{\text{T}}/n_{\text{D}}$ [14]. Of particular relevance in SPARC is the $^3$He concentration $n_{\text{He}-3}/n_{\text{e}}$, since $^3$He will be used for minority heating at its ion cyclotron resonance frequency. Hence, CTS could facilitate core-localized measurements of the non-negligible $^3$He concentration in SPARC.

Previous work has highlighted the diagnostic potential of CTS in compact, high-field tokamaks [15]. This was based on predicted scattering functions using an electrostatic model of CTS and assuming specific scattering geometries. The results showed that the CTS signal of fusion-born alphas in SPARC will clearly dominate the signal of the bulk plasma at certain frequency ranges of the CTS spectrum, thus potentially providing useful data on the $\alpha$ density in the plasma core.

Here we extend the work of [15] by identifying realizable scattering geometries in SPARC, based on the tokamak layout, magnetic equilibrium, and kinetic plasma profiles of SPARC. We predict CTS spectra using a fully electromagnetic scattering model, along with diagnostic signal-to-noise ratios based on realistic background noise estimates. Furthermore, we assess the capability of a CTS system to measure relevant plasma parameters using "off-the-shelf" gyrotrons with operating frequencies already in use at existing devices. Given the short timeline for SPARC construction, with initial machine commissioning planned for 2025 [1], the use of standard gyrotron frequencies is an important element in the timely development of a CTS diagnostic for the operational phases of SPARC.

This paper is organized as follows: section 2 presents the general principles of CTS and the SPARC plasma scenario underlying our analysis. Section 3 outlines a conceptual design and technical specifications of a CTS system for SPARC, including probe frequency and polarization, viable scattering geometries, and diagnostic background noise. Resulting synthetic CTS spectra and signal-to-noise ratios are presented in section 4, along with a brief sensitivity study of relevant plasma parameters. In section 5, we explore alternative scattering geometries, followed by a general discussion of the diagnostic design in section 6. Finally, section 7 summarizes our key findings and outlines directions for future work.

## 2. Methods and assumptions

### *2.1. The principles of CTS*

CTS is based on injecting powerful electromagnetic waves into the plasma and collecting a fraction of the resulting scattered radiation. The scattered signal will be dominated by signatures of ion motion, provided that the Salpeter parameter,

$$\alpha_{\mathrm{SP}} = \frac{1}{k^\delta \lambda_{\mathrm{D}}}, \quad (1)$$

is well above 1. Here $\mathbf{k}^\delta = \mathbf{k}^{\mathrm{s}} - \mathbf{k}^{\mathrm{i}}$ is the wave vector of the plasma fluctuations resolved by CTS, $\mathbf{k}^{\mathrm{i}}$ and $\mathbf{k}^{\mathrm{s}}$ are the wave vectors of the incident (probing) and scattered waves, respectively, and $\lambda_{\mathrm{D}}$ is the electron Debye length in the scattering volume (SV). Formally, the SV is here defined as the volume within which 75% of the scattered signal originates.

The resulting scattering spectrum is sensitive to the 1D ion velocity distribution function $g(u)$, which is a projection of the full ion velocity distribution function $f(\mathbf{v})$ onto $\mathbf{k}^\delta$,

$$g(u) = \int f(\mathbf{v}) \delta\left(\frac{\mathbf{v} \cdot \mathbf{k}^\delta}{k^\delta} - u\right) \mathrm{d}\mathbf{v}, \quad (2)$$

where $\delta$ is the Dirac $\delta$–function, $\mathbf{v} \cdot \mathbf{k}^\delta / k^\delta \approx 2\pi \Delta v / k^\delta$ is the ion velocity projected along $\mathbf{k}^\delta$, and $\Delta v = v - v_{\mathrm{gyr}}$ is the frequency shift relative to the probe frequency $v_{\mathrm{gyr}}$ at which an ion with velocity $\mathbf{v}$ contributes to the scattered spectrum. The CTS spectrum therefore carries information about the thermal- and fast-ion velocity distribution, including the thermal-ion temperature $T_i$, and the plasma toroidal rotation velocity $v_{tor}$. For scattering geometries with angles $\phi = \angle(\mathbf{k}^\delta, \mathbf{B}) \approx 90°$ between the resolved fluctuation wave vector $\mathbf{k}^\delta$ and the magnetic field $\mathbf{B}$, the CTS spectrum contains signatures of ion cyclotron motion and Bernstein waves associated with the various ion species in the plasma [16]. Such scattering geometries are sensitive to the plasma composition in the SV, facilitating measurements of the local fuel-ion ratio $R_i$ [17] for example.

The scattered spectral power density (SPD) $\partial P^s / \partial \omega^s$ at a given angular frequency $\omega^s$ is described by the CTS transfer equation,

$$\frac{\partial P^s}{\partial \omega^s} = P^i O_b n_e \lambda_0^i \lambda_0^s r_e^2 \Sigma, \qquad (3)$$

where $P^i$ is the incident probe power in the SV, $O_b$ is the overlap factor quantifying the geometric overlap between the incident and received beams, $\lambda_0^i$ and $\lambda_0^s$ are the incident and scattered vacuum wavelengths, respectively, $r_e$ is the classical electron radius, and $\Sigma$ is the scattering function [18].

The dominant contribution to the diagnostic background noise for existing CTS diagnostics stems from electron cyclotron emission (ECE). The noise is generally quantified by the standard deviation of the estimated post-integration CTS signal [19],

$$\sigma_{CTS} = \sqrt{\frac{2(P^s + P^b)^2 + 2(P^b)^2}{\Delta f_b \Delta t}}, \qquad (4)$$

where $P^s$ and $P^b$ is the spectral power density of the CTS signal and background noise, respectively, falling within a specific frequency channel of width $\Delta f_b$, and $\Delta t$ is the total useful integration time. The signal-to-noise ratio is then given by $S/N = P^s / \sigma_{CTS}$.

Designing a CTS diagnostic thus partly revolves around optimizing $\alpha_{SP}$ and $O_b$ in eqs. (1) and (3), respectively, while minimizing the noise in eq. (4) through a judicious choice of operating frequency, scattering geometry, and polarization mode for plasma accessibility.

*2.2. Plasma scenario*

In order to predict realizable CTS scattering geometries and resulting diagnostic signal and background levels, we obtained the magnetic equilibrium for the SPARC double null primary reference discharge (PRD) [20] from the SPARC Public GitHub repository [21]. The resulting magnetic flux contours and $\mathbf{B}$-field are shown in fig. 1(a) and (b). Associated profiles of electron density $n_e$ and electron (ion) temperature $T_e$ ($T_i$), are shown in fig. 1(c) as functions of the normalized poloidal flux coordinate $\rho_p = \sqrt{\psi_N}$. These profiles are adopted from the core plasma profiles presented in [20, 22], and available from [21]. For the purpose of raytracing of the diagnostic beam paths, $T_e$ and $n_e$ have here been extrapolated beyond the last closed flux surface (LCFS, $\rho_p = 1$) using an inverse hyperbolic tangent scheme.

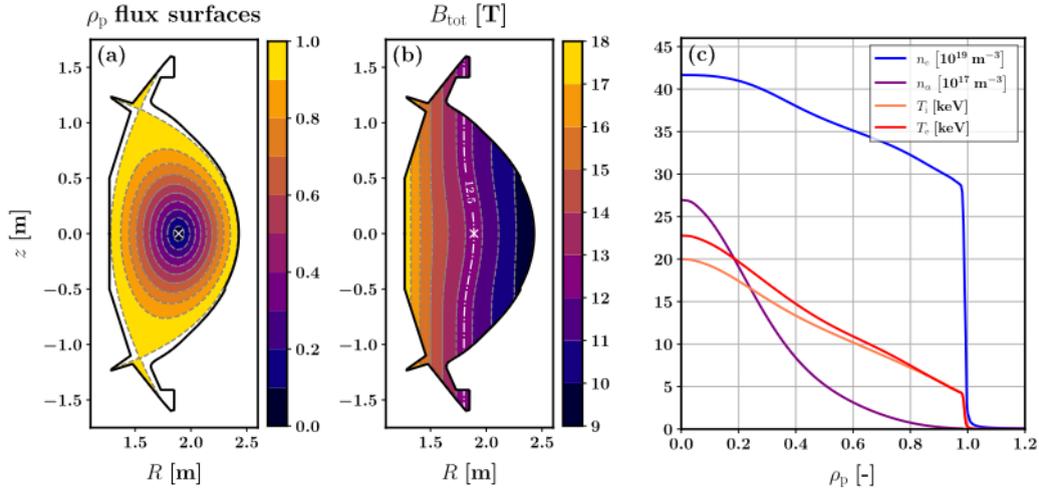

**Figure 1.** Magnetic equilibrium and plasma profiles for the SPARC PRD utilized in this study, adopted from [21]. (a) Contours of the normalized poloidal flux $\rho_p = \sqrt{\psi_N}$, with a poloidal cross section of the machine layout shown in black and '×' indicating the magnetic axis. (b) Contours of the total magnetic field $B_{tot}$ with on-axis $B_0 \approx 12.5$ T. (c) Electron density $n_e$ and temperature $T_e$, and ion temperature $T_i$, with $n_e$ and $T_e$ extrapolated beyond the LCFS as described in the main text. The fusion-born $\alpha$ density $n_\alpha$ was estimated using an isotropic slowing-down velocity distribution.

All ion species present in the plasma contribute to the total CTS signal. Here we incorporate thermalized D and T as majority ion species, with densities $n_D = n_T = 0.425 n_e$, along with an assumed W impurity density $n_W = 1.5 \times 10^{-5} n_e$ with charge $Z \approx +51$ [20, 22]. Also included is a contribution from thermalized $^4$He ash with $n_{He-4} = 0.015 n_e$, based on a volume-averaged estimate at the end of the flat-top phase [20]. Furthermore, we assume the presence of minority $^3$He for use with ICRH at a level of $n_{He-3} = 0.05 n_e$ [20], along with $n_H = 0.01 n_e$ from D-D fusion in order to help satisfy global charge neutrality. Non-thermalized ions in the plasma include fusion-born alphas, see fig. 1(c), for which we assume an isotropic slowing-down velocity distribution [23, 24], with the fusion reaction rate estimated using [25].

Lastly, SPARC will not apply neutral beam injection and so will feature no external angular momentum input. This might suggest low levels of toroidal bulk plasma rotation, i.e. $v_{tor} \approx 0$, but empirical predictions indicate some intrinsic plasma rotation in SPARC at the level of a core Mach number $\mathcal{M} \sim 0.16$ [22, 26]. This would correspond to a toroidal bulk-ion core rotation of $v_{tor} \sim 200 \, \text{km s}^{-1}$ for an estimated on-axis ion sound speed $\sim 1250 \, \text{km s}^{-1}$ in a pure D-T plasma. In our analysis of the CTS spectra, we will assume $v_{\text{tor}} = 0$ for the CTS geometries that are at most marginally sensitive to plasma rotation, and $v_{\text{tor}} = 200\,\text{km}\,\text{s}^{-1}$ for all others.

## 3. SPARC CTS pre-conceptual design

### 3.1. Probe frequency and polarization

Any SPARC CTS diagnostic measuring in the plasma core will need to rely on backwards scattering, since the compact design and dual divertors of SPARC do not allow diagnostic infrastructure located on the high-field side of the tokamak. This restricts the operating frequency of the diagnostic to the microwave range below the SPARC electron cyclotron resonance in order to satisfy the Salpeter criterion of $\alpha_{\text{SP}} > 1$.

Figure 2 illustrates the relevant plasma cutoff and resonance frequencies in SPARC based on the magnetic equilibrium and electron profiles of fig. 1. With a core plasma frequency $f_{\text{pe}} \approx 180\,\text{GHz}$, O-mode CTS operation would realistically require probing frequencies $f \gtrsim 200\,\text{GHz}$. However, as shown in section 3.3, the diagnostic background noise at such frequencies becomes significant without any corresponding benefit to the CTS signal level. In addition, the results of [15] suggest that using frequencies $f \lesssim 200\,\text{GHz}$ would be beneficial for $\alpha$-particle measurements in SPARC. Hence, X-mode operation is the obvious choice for a CTS diagnostic in SPARC when capability to measure in the plasma core is required.

From a technological readiness perspective, CTS frequencies of 105, 140, or 175 GHz would be attractive options, since gyrotrons operating at, or close to, these frequencies are already available at existing devices including DIII-D [27], ASDEX Upgrade [28, 29], and Wendelstein 7-X [30]. These frequencies lie well below the fundamental electron cyclotron resonance frequency $f_{\text{ce}} \gtrsim 240\,\text{GHz}$ in SPARC, while also avoiding both the X-mode L-cutoff ($f_{\text{L}} \lesssim 80\,\text{GHz}$) and the upper hybrid resonance frequency throughout the plasma. The latter point is important for preventing parametric decay of the probe beam and associated daughter waves affecting the resulting CTS spectrum [31, 32, 33]. Within the 105–175 GHz frequency range, the exact choice of frequency represents a compromise between increased ECE noise at higher frequencies and increased refraction of the diagnostic beams at lower frequencies approaching $f_{\text{L}}$.

### 3.2. Scattering geometries

SPARC will be equipped with 18 toroidally distributed sets of three ports for, e.g., diagnostics and heating systems. For each set, one is an equatorial port centered at $R \approx 2.45\,\text{m}$ and $z \approx 0\,\text{m}$, while the other two are located at $R \approx 2.15\,\text{m}$ and are vertically staggered into upper/lower ($z \approx \pm 0.75\,\text{m}$) ports [2]. In the following, we assume the CTS probe to be located in an upper port and the receiver in the corresponding lower port. Both probe and receiver beams are modelled as Gaussian beams, with parameters based on those of the electron cyclotron heating beams in ASDEX Upgrade [34, 35], whose vessel dimensions are comparable to those of SPARC. Table 1 lists these parameters, with the 105 and 140 GHz values identified from in-vessel alignment measurements and then linearly extrapolated to 175 GHz. Additionally, in section section 5 we investigate the implications of a CTS system located exclusively in an equatorial port.

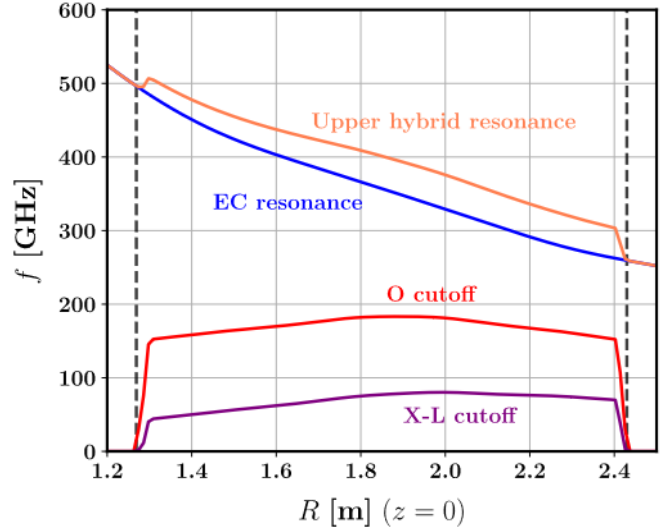

**Figure 2.** SPARC PRD cutoff frequencies for O-mode (electron plasma frequency $f_{pe}$) and X-mode (L-cutoff $f_L$) waves as a function of major radius coordinate $R$ in the midplane ($z = 0$). Also shown are the corresponding upper hybrid, $f_{UH}$, and electron cyclotron, $f_{ce}$, resonance frequencies. Dashed vertical lines represent the vacuum vessel walls. All cutoffs and resonances are calculated in the cold plasma approximation.

**Table 1.** Adopted Gaussian beam radius $w_0$ at the beam waist and waist distance $s_0$ from the launcher mirrors for the considered frequencies $f$.

| $f$ | 105 GHz | 140 GHz | 175 GHz |
|---|---|---|---|
| $w_0$ | 22.9 mm | 17.0 mm | 11.1 mm |
| $s_0$ | 489.1 mm | 477.1 mm | 465.1 mm |

Using raytracing with our in-house code *Warmray* [18, 19, 36], we have performed a sweep of toroidal and poloidal injection angles of both probe and receiver beams to determine scattering geometries that facilitate useful CTS measurements in the plasma core. We define the poloidal angle $\alpha$ as the injection angle relative to horizontal in the poloidal cross section (counted positive downwards) and the toroidal angle $\beta$ as the injection angle relative to the radial direction in the toroidal cross section. For simplicity, we have considered only symmetrical configurations, with the probe launching angles $(\alpha,\beta)$ and the corresponding receiver angles $(-\alpha,\beta)$. Figure 3 shows the results in terms of the calculated values of relevant CTS parameters.

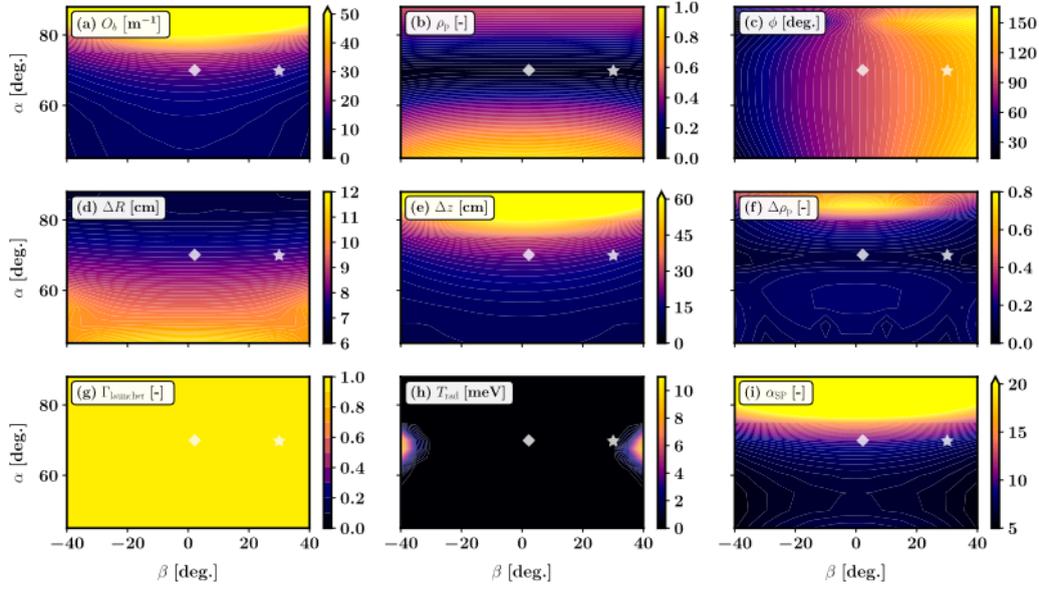

**Figure 3.** Characteristic parameters of the CTS scattering geometries as a function of poloidal $\alpha$ and toroidal $\beta$ launching angles of the probe $(\alpha,\beta)$ and receiver $(-\alpha,\beta)$ beam for 140 GHz X-mode. The probe is in an upper port and the receiver in the corresponding lower port. (a) Overlap factor $O_b$. (b) Poloidal flux coordinate $\rho_p$ at the SV center. (c) CTS measurement angle $\phi = \angle(\mathbf{k}^\delta, \mathbf{B})$. (d)–(f) Extent of the SV along $R$, $z$ and $\rho_p$, respectively. (g) Transmittance $\Gamma$ of the central probe ray along the path to the SV. (h) Single-pass radiation temperature increment $T_{\text{rad}}$ along the receiver view. (i) Salpeter parameter $\alpha_{\text{SP}}$. The ◆ and ★ represent the launching angles of the perpendicular and the non-perpendicular geometries of table 2 at 140 GHz, respectively.

Suitable scattering geometries were selected based on maximizing the overlap factor $O_b$ between probe and receiver beams as well as the Salpeter parameter $\alpha_{\text{SP}}$, while simultaneously minimizing $\rho_p$ (to center the SV in the plasma core) and $\Delta\rho_p$. As indicated by fig. 3, the requirement of measuring in the plasma core provides a strong constraint on the possible range of many other parameters. Improved spatial resolution is primarily achieved by decreasing $\Delta\rho_p$ rather than $\Delta R$ and $\Delta z$ due to the geometry of the flux surfaces. In addition, we attempted either to maximize $|\phi - 90°|$ in order to obtain geometries with measurement angles non-perpendicular to the magnetic field for good sensitivity to plasma rotation $v_{\text{tor}}$ and fast ions, or ensure a perpendicular geometry with $|\phi - 90°| \lesssim 3°$ for good sensitivity to plasma composition such as the fuel-ion ratio $R_i$.

Figure 4 illustrates examples of raytracing results for the identified "optimum" geometries of both types, and table 2 summarizes the salient parameters for each identified geometry at the relevant frequencies. The conditions in SPARC evidently allow core-localized ($\rho_p < 0.06$) CTS measurement volumes with good overlap factors $O_b$, large values of $\alpha_{SP}$, and decent spatial resolution with $\Delta \rho_p < 0.13$ and $\Delta R, \Delta z \in [8, 27]$ cm.

**Table 2.** CTS parameters for identified "optimum" geometries.

| $f$ | 105 GHz | 140 GHz | 175 GHz |
|---|---|---|---|
| Non-perpendicular geometries | | | |
| $(\alpha, \beta)$ | (67.5°, 25°) | (70°, 30°) | (70°, 30°) |
| $\phi$ | 127.4° | 127.6° | 125.0° |
| $O_b$ [m$^{-1}$] | 8.4 | 14.5 | 15.6 |
| $\alpha_{SP}$ | 12.6 | 8.6 | 6.5 |
| $\rho_p$ | 0.05 | 0.05 | 0.06 |
| $\Delta R$ [m] | 0.12 | 0.08 | 0.08 |
| $\Delta z$ [m] | 0.20 | 0.18 | 0.19 |
| $\Delta \rho_p$ | 0.10 | 0.09 | 0.09 |
| Perpendicular geometries | | | |
| $(\alpha, \beta)$ | (67.5°, 2°) | (70°, 2°) | (70°, 2°) |
| $\phi$ | 92.9° | 92.5° | 92.3° |
| $O_b$ [m$^{-1}$] | 11.5 | 17.6 | 17.5 |
| $\alpha_{SP}$ | 13.7 | 9.9 | 7.3 |
| $\rho_p$ | 0.06 | 0.05 | 0.06 |
| $\Delta R$ [m] | 0.12 | 0.08 | 0.08 |
| $\Delta z$ [m] | 0.27 | 0.21 | 0.21 |
| $\Delta \rho_p$ | 0.13 | 0.10 | 0.10 |

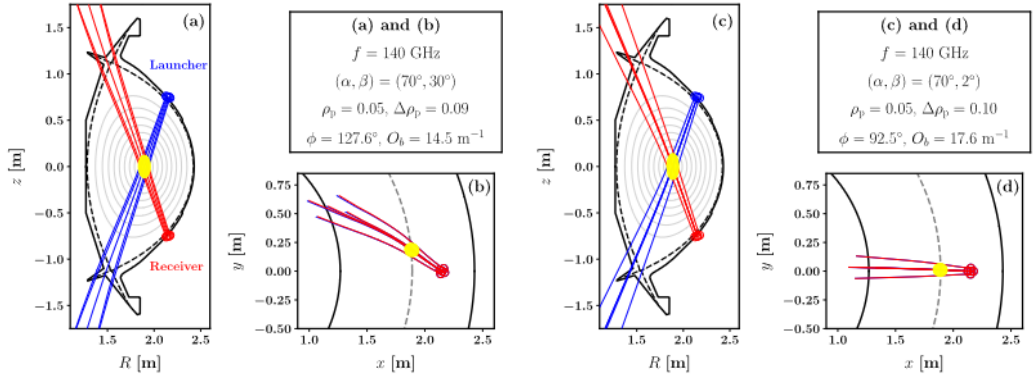

**Figure 4.** "Optimum" scattering geometries for core CTS measurements at 140 GHz in poloidal and toroidal cross sections of SPARC for the (a,b) non-perpendicular and (c,d) perpendicular geometries listed in table 2. For (a) and (c), the $\rho_p$ flux surfaces from fig. 1(a) are shown in grey, along with the LCFS in dashed black. For (b) and (d), black lines represent the vessel walls at $z = 0$, and dashed grey line marks the magnetic axis. The intersection of probe rays (blue) and receiver rays (red) represents the SV shown in yellow.

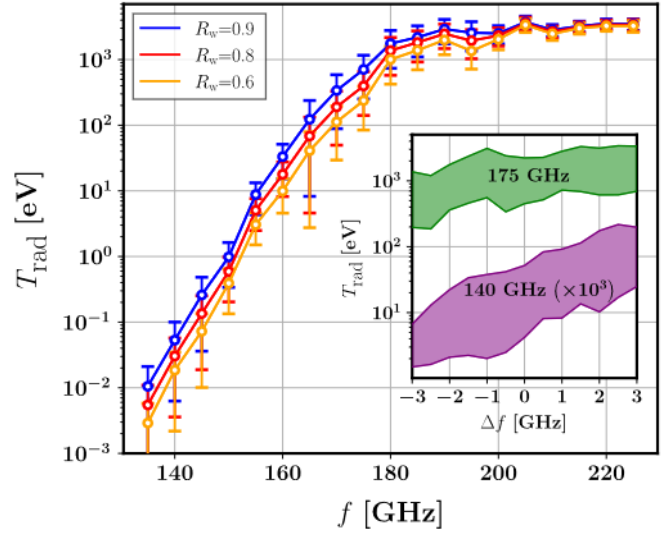

**Figure 5.** Predicted X-mode ECE radiation temperature $T_{\rm rad}$ in SPARC for the PRD plasma scenario (section 2.2) across relevant frequencies $f$ for different wall reflectivities $R_{\rm w}$. The statistical error at each $f$ stems from the Monte Carlo method described in the main text. The inset shows $T_{\rm rad}$ at a given frequency shift $\Delta f$ from relevant probe gyrotron frequencies. Shaded regions encompass the statistical uncertainties from the Monte Carlo-based procedure and a variation associated with assuming different $R_{\rm w} \in [0.6, 0.9]$. Results at 140 GHz have been increased by a factor of $10^3$ for ease of comparison.

*3.3. Diagnostic background noise*

For existing CTS diagnostics, ECE dominates the diagnostic background noise. However, the probe frequencies considered here all lie considerably below the fundamental electron cyclotron resonance frequencies in SPARC (section 3.1). In this regime, the plasma is optically thin, and standard single-pass ECE calculations are unreliable for assessing the ECE signal level and its frequency dependence. In order to estimate ECE noise, and hence CTS signal-to-noise ratios, multiple ECE ray passes subject to vessel wall reflections must be considered.

To this end, we applied a Monte Carlo procedure originally developed for the ITER CTS diagnostic [37] and benchmarked against results from the SPECE raytracing code [38]. Using *Warmray*, a ray is traced backwards from the CTS receiver through hundreds of diffuse, randomized wall reflections in the vessel wall geometry. From this, we calculate the ECE radiation temperature $T_{\rm rad}$ observed by the receiver, including the effects of wall absorption and polarization scrambling for each wall reflection. For sufficiently diffuse reflections, the viewing geometry of the receiver has no bearing on the results [37, 39], and the only free parameter in the approach is the adopted wall reflectivity $R_{\rm w}$ at the relevant frequencies.

**Table 3.** Estimated contributions to the total CTS background noise at different probe frequencies $f$.

| $f$ | 105 GHz | 140 GHz | 175 GHz |
|---|---|---|---|
| ECE noise ($T_{\text{rad}}$) | 0 eV | 1 eV | 1 keV |
| Non-ECE noise | 1 eV | 1 eV | 1 eV |
| Total noise ($P^{\text{b}}$) | 1 eV | 2 eV | $\approx$ 1 keV |

Figure 5 shows the results of this procedure for the X-mode ECE radiation temperature across a plausible range in $R_{\text{w}}$, using the wall geometry and plasma equilibrium of fig. 1. A strong frequency dependency of $T_{\text{rad}}$ is observed across the considered interval $f \in [135, 220]$ GHz. However, for $f \gtrsim 180$ GHz, the slope of $T_{\text{rad}}(f)$ decreases as $T_{\text{rad}}$ slowly approaches the core $T_{\text{e}}$. At these frequencies, the predicted ECE background can reach $T_{\text{rad}} \gtrsim 1$ keV. This represents a strong argument in favour of adopting a lower CTS probe frequency for SPARC. In contrast, the background levels at frequencies $f \lesssim 140$ GHz are almost negligible, $T_{\text{rad}} \ll 1$ eV.

Across a typical frequency width of a measured CTS spectrum of 5–10 GHz, the systematic increase in $T_{\text{rad}}$ with frequency implies that the $S/N$ will be lower for the upshifted parts of the CTS spectrum. This effect is relatively more pronounced at 140 GHz compared to at 175 GHz. Nevertheless, a single value for $T_{\text{rad}}$ can conservatively be applied to assess the $S/N$ across the entire CTS spectrum around a given probe frequency. Based on fig. 5, we assume a vanishing $T_{\text{rad}} \approx 0$ eV at 105 GHz, while $T_{\text{rad}} = 1$ eV represents a conservative choice at 140 GHz. Similarly, $T_{\text{rad}} = 1$ keV was chosen as a representative, but slightly less conservative value, at 175 GHz.

Non-ECE noise for the diagnostic consists of the internal receiver noise, transmission line loss, and potentially other external contributions. Here we adopt a uniform non-ECE noise level of approximately 1 eV for all considered probe frequencies, based on experience from the CTS diagnostic at ASDEX Upgrade [40]. Conservatively assuming that the ECE and non-ECE noise levels add linearly, we summarize the results in table 3. For the 105 and 140 GHz options, the total diagnostic noise $P^{\text{b}}$ has a significant contribution from receiver noise and, e.g., calibration uncertainties.

## 4. Predicted CTS measurements

### 4.1. Synthetic spectra and signal-to-noise ratios

Based on the identified scattering geometries (table 2), and on the plasma profiles from section 2.2, we generated synthetic CTS spectra for each geometry using a fully electromagnetic scattering model [41]. The results, shown in fig. 6, assume an injected probe power of $P^i = 200\,\text{kW}$. This relatively low value of $P^i$ was adopted to reduce the predicted wall heat load from the largely unabsorbed CTS probe gyrotron, and to facilitate integration of the diagnostic within the overall diagnostic power budget for SPARC. The calculated signal-to-noise ratios assume the noise levels of table 3, along with a frequency resolution of $\Delta f_b = 5\,\text{MHz}$ and integration time of $\Delta t = 10\,\text{ms}$. Furthermore, a notch filter in the CTS receiver electronics must be present to protect against stray radiation from the probe gyrotron. Here we assume a width of this notch filter of 250 MHz, similar to that implemented for the CTS diagnostic at ASDEX Upgrade [42].

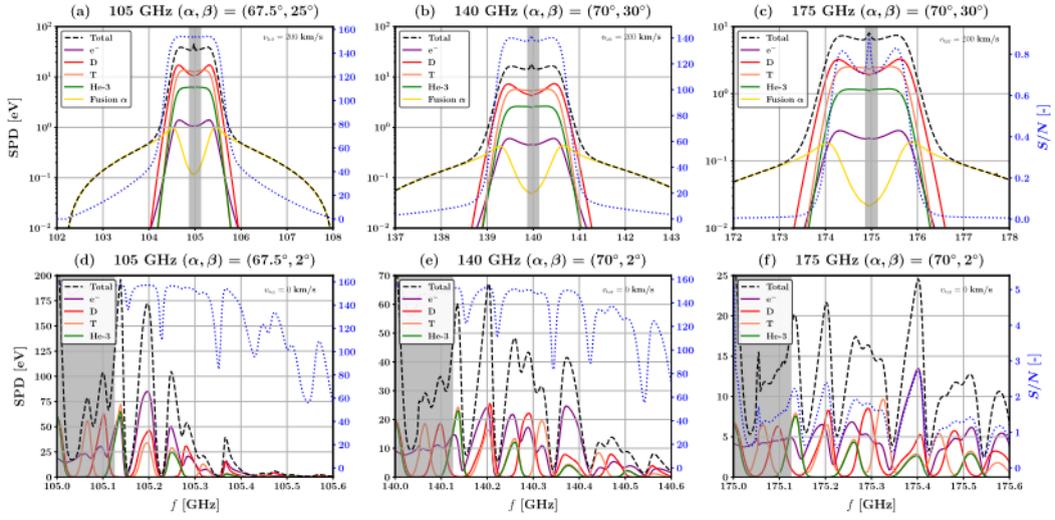

**Figure 6.** Synthetic CTS spectra for the (top) non-perpendicular scattering geometries and (bottom) perpendicular geometries from table 2. The total SPD is shown in dashed black, along with the contributions from electrons, D, T, $^3$He, and fusion-born alphas. The values (SPD) of these signals should be read on the left vertical axis. Note that for (a)–(c) the left vertical axis is logarithmic, and for (d)–(f) the left vertical axis is linear. The corresponding signal-to-noise ratio of the total CTS signal (dotted blue) is indicated on the right (linear) vertical axis. The grey region shows the frequency interval affected by the assumed notch filter. For improved visibility, only the upshifted part of the spectra are shown in (d)–(f). The spectra were calculated assuming (a)–(c) $v_{\text{tor}} = 200\,\text{km s}^{-1}$ and (d)–(f) $v_{\text{tor}} = 0$.

In all cases, the predicted total ion signal clearly dominates over the electron contribution, as expected for the values of $\alpha_{\text{SP}}$ in table 2. For the non-perpendicular geometries, the characteristic width of the bulk signal clearly increases with probe frequency across the explored frequency range. In contrast, the typical signal level decreases with increasing $f_{\text{gyr}}$ in all cases, in part due to the associated decrease in wavelength, see eq. (3). Additionally, the variation in $O_b$ (cf. table 2) linearly affects the typical SPD, along with, to a lesser extent, the variation of $\alpha_{\text{SP}}$.

For the non-perpendicular geometries, the peak total SPD ranges from 5–40 eV for the bulk-ion feature outside the notch filter range. This is easily a detectable signal level, and for 105 and 140 GHz, $S/N > 10$ for all frequencies between the bulk signal and until the $\alpha$ signal dominates. However, for 175 GHz $S/N \ll 10$ at all frequencies. As such, reliable measurements of, e.g., $T_i$ would be feasible only at 105 and 140 GHz for the assumed combination of $\Delta f_b \Delta t$.

We assume $v_{\text{tor}} = 200\,\text{km}\,\text{s}^{-1}$ (see section 2.2) for the non-perpendicular geometries. This gives rise to a visible frequency downshift of the entire spectrum for the probe frequencies in fig. 6. For these geometries, this is attributed to $|\phi - 90°| \gg 0$, providing decent sensitivity to bulk ion rotation along **B**. For the perpendicular geometries ($\phi \approx 90°$), the spectral downshift would be challenging to identify, so we simply adopted $v_{\text{tor}} = 0$ for these as explained in section 2.2.

The assumed frequency resolution of $\Delta f_b = 5\,\text{MHz}$ is clearly sufficient to adequately represent the spectra. For the non-perpendicular geometries, it would be feasible, and even recommendable, to increase $\Delta f_b$ and hence improve $S/N$ while maintaining sufficient spectral resolution. However, the perpendicular geometries impose strict requirements on $\Delta f_b$ in order to properly sample the wavelike structures of the CTS signal. Based on the Nyquist-Shannon sampling theorem, a rough requirement of $\Delta f_b \lesssim 30\,\text{MHz}$ would be imposed by the need to properly sample the contribution from T, the ion species with the smallest cyclotron frequency among the considered ion species.

For the 105 and 140 GHz perpendicular geometries, $S/N > 50$ even at frequency shifts $|\Delta f| = 0.6\,\text{GHz}$ from the probe frequency. Clearly, D, T and $^3$He all contribute noticeably to the total signal. This underscores the possibility of fuel-ion ratio and $^3$He measurements at these frequencies. However, the 175 GHz perpendicular geometry appears unattractive for this purpose due to the low $S/N$.

*4.2. Prospects for fusion-born $\alpha$ measurements*

As established in [15], the fraction of the entire CTS spectrum dominated by the $\alpha$ signal increases for decreasing probe frequencies $f$, potentially allowing better measurement constraints on the $\alpha$ contribution at lower $f$. However, the characteristic width of the spectra also decreases with decreasing $f$ (fig. 6(a-c)), thus reducing the relevant absolute frequency range. While these considerations are important, they must be accompanied by corresponding signal-to-noise estimates to assess the capability of our non-perpendicular geometries to determine the fusion-born $\alpha$ distribution function.

For this purpose, we introduce $f_{90\%,l}$ and $f_{90\%,u}$ as the frequencies below and above which the $\alpha$ signal comprises at least 90% of the total signal. Table 4 lists these for each probe frequency. For frequencies $f < f_{90\%,l}$, the $\alpha$ signal is thus nearly identical to the total signal until the electron contribution becomes dominant at lower $f$ (and correspondingly for $f > f_{90\%,u}$). Next, we define $S/N(90\% \pm 1\,\text{GHz})$ as the signal-to-noise ratio averaged over the union of the two frequency intervals $[f_{90\%,\text{l}} - 1\,\text{GHz}, f_{90\%,\text{l}}]$ and $[f_{90\%,\text{u}}, f_{90\%,\text{u}} + 1\,\text{GHz}]$. This provides an estimate of the effective $S/N$ of the total CTS signal across a measurement-relevant fraction of the $\alpha$-dominated frequency range.

**Table 4.** The lower and upper frequencies $f_{90\%,\text{l}}$ and $f_{90\%,\text{u}}$ at which the $\alpha$ signal first contributes 90% of the total CTS signal for the non-perpendicular geometries.

| $f$ | 105 GHz | 140 GHz | 175 GHz |
|---|---|---|---|
| $f_{90\%,\text{l}}$ | 104.08 MHz | 138.66 MHz | 173.23 MHz |
| $f_{90\%,\text{u}}$ | 105.89 MHz | 141.27 MHz | 176.68 MHz |

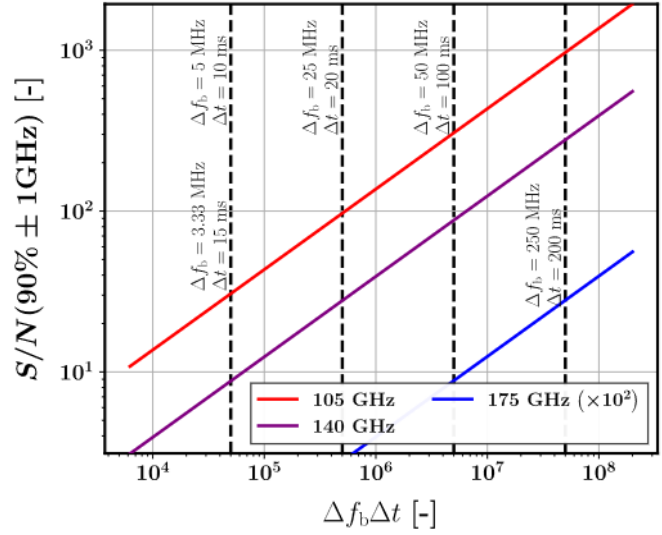

**Figure 7.** Signal-to-noise ratio $S/N$ of the total CTS signal averaged over 2 GHz of the $\alpha$-dominated frequency range as a function of the product of frequency resolution $\Delta f_\text{b}$ and integration time $\Delta t$. The three cases shown are the non-perpendicular geometries from table 2. The dashed black isolines highlight different combinations of $\Delta f_\text{b}$ and $\Delta t$ (the leftmost isoline corresponds to the case from fig. 6). Results for 175 GHz have been increased by a factor of $10^2$ to improve visibility.

As shown in fig. 7, this averaged $S/N$ is well above 10 at 105 GHz for $\Delta f_\text{b} = 5\,\text{MHz}$ and $\Delta t = 10\,\text{ms}$. A similar result can be obtained at 140 GHz for $\Delta f_\text{b} = 25\,\text{MHz}$ and $\Delta t = 20\,\text{ms}$. These frequency resolutions are more than adequate (cf. section 4.1), and the values of $\Delta t$ are well below the estimated fusion-born $\alpha$ slowing-down time in the plasma core of $t_{\text{SD},\alpha} \sim 200\,\text{ms}$. At 175 GHz, however, the similarly averaged $S/N$ remains below unity, even for $\Delta f_\text{b} = 250\,\text{MHz}$ and a time resolution approaching that of $t_{\text{SD},\alpha}$.

With typical $\alpha$ signal levels of 0.1–0.5 eV for 105 and 140 GHz, characterization of this component in CTS spectra will require good control of systematic noise contributions. Nevertheless, these signal levels are comparable to those anticipated from the ITER CTS system and are accompanied by lower predicted ECE levels [37, 43]. As such, we expect the accuracy of CTS-based fusion-born alpha density measurements in SPARC to be at least comparable to that obtainable at ITER, i.e., $\sim 10\%$ at a temporal resolution of $\Delta t = 20\,\mathrm{ms}$ [43]. These results demonstrate the promising prospect of measuring the core-localized fusion-born $\alpha$ distribution in SPARC at both 105 and 140 GHz.

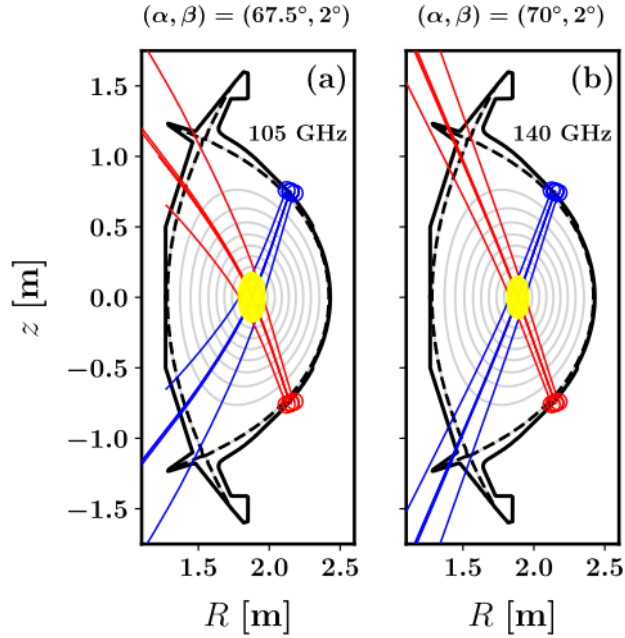

**Figure 8.** Raytracing results for the perpendicular scattering geometries from table 2 at (a) 105 GHz and (b) 140 GHz in a SPARC poloidal cross section. The case in (b) is identical to that in fig. 4(c) and is repeated here for ease of comparison.

*4.3. Sensitivity to refraction*

The results in sections 4.1 and 4.2 indicate that useful signal-to-noise ratios can be obtained for a CTS diagnostic at SPARC operating at 105 or 140 GHz. A significant difference between these two options is the impact of refraction on the scattering geometry, with 105 GHz lying considerably closer to the L-cutoff (section 3.1). Based on table 2, refraction clearly does not prevent centralized SVs at 105 GHz, although this is accompanied by larger SVs and hence reduced spatial resolution compared to the 140 GHz and 175 GHz cases. Specifically, for the 140 GHz perpendicular geometry, the volume of the SV is $\approx 608\,\mathrm{cm}^3$ while for the 105 GHz analog, the volume is $\approx 1540\,\mathrm{cm}^3$.

For the perpendicular geometries, $\phi$ must remain stable around 90° to allow plasma composition measurements. Susceptibility to refraction could here pose a significant issue, in particular under fluctuating plasma conditions such as those driven by instabilities and turbulence. Figure 8 illustrates the 105 and 140 GHz perpendicular geometries from table 2. While refraction effects are small at 140 GHz, they lead to bending and broadening of the diagnostic beam at 105 GHz, resulting in a larger SV. Additionally, the larger intrinsic beamwidth at 105 GHz coupled with refractive broadening results in a lower beam overlap $O_b$, as shown in table 2. For large toroidal angles $\beta$, refraction becomes noticeable in toroidal cross sections at 140 GHz as well.

In contrast, refraction is largely insignificant for all raytraces at 175 GHz. However, the generally low $S/N$ at 175 GHz seen in figs. 6 and 7 is not outweighed by any other advantage at this probing frequency, so 175 GHz is a less attractive option. With a judicious choice of $\Delta f_b \Delta t$, both 105 and 140 GHz provide ample $S/N$, but the enhanced sensitivity to refraction at 105 GHz could be a critical challenge for isotope measurements. This leaves 140 GHz as our recommended option for a general-purpose CTS diagnostic at SPARC.

*4.4. Bulk-ion measurement capabilities*

To assess the bulk-ion measurement capabilities of the CTS diagnostic, we consider the sensitivity of the total CTS signal to variations in thermal-ion plasma parameters.

For the non-perpendicular 140 GHz geometry, varying $v_{\text{tor}}$ is instructive. Defining $\Delta\text{SPD}(v_{\text{tor}})$ as the difference in total signal between a plasma with toroidal rotation $v_{\text{tor}}$ and a plasma with no rotation, we obtain the results in fig. 9(a). For $v_{\text{tor}} = 200\,\text{km}\,\text{s}^{-1}$, peaks in $\Delta\text{SPD}(v_{\text{tor}})$ of up to 2 eV are present outside the notch filter range. These values are comparable to the assumed noise level of $\approx 2\,\text{eV}$ at 140 GHz, and the peaks persist across several hundred MHz. Thus, we expect a signal difference between a non-rotating plasma and a plasma with $v_{\text{tor}} = 200\,\text{km}\,\text{s}^{-1}$ to be resolvable with this 140 GHz scattering geometry.

As previously stated, measurements of $T_i$ with the 140 GHz non-perpendicular geometry also seem feasible. To support this assertion, we vary the ion temperature $T_{i,\text{SV}}$ in the SV and quantify the signal difference $\Delta\text{SPD}(k_{T_i})$ compared to that of the baseline $T_{i,\text{SV}} \approx 19.8\,\text{keV}$ from the $T_i$ profile of section 2.2 at the SV center. Here $k_{T_i}$ is a scaling factor on $T_{i,\text{SV}}$. From fig. 9(b), we see that $k_{T_i} = 1.25$, corresponding to $T_{i,\text{SV}} \approx 24.8\,\text{keV}$, yields peaks of up to 2 eV in $\Delta\text{SPD}(k_{T_i})$. Reusing the above argument, bulk-ion temperature changes of at least 25% should lead to detectable changes in the total signal difference.

For the perpendicular 140 GHz geometry, we have varied the fuel-ion ratio $R_i$ while maintaining quasi-neutrality in the SV. Figure 9(c) shows the resulting variation in signal $\Delta\text{SPD}(R_i)$ relative to the baseline case of $R_i = 1$ (50:50 D-T mixture) with $n_D = n_T = 0.425 n_e$ (section 2.2). For $R_i = 1.22$ (45:55 D-T mixture), corresponding to a 10% decrease in $n_D$ and a 10% increase $n_T$, narrow peaks of several eV appear just outside the notch filter range. As such, changes of $\approx 10\%$ in $n_D$ and $n_T$ should result in a measurable total signal difference.

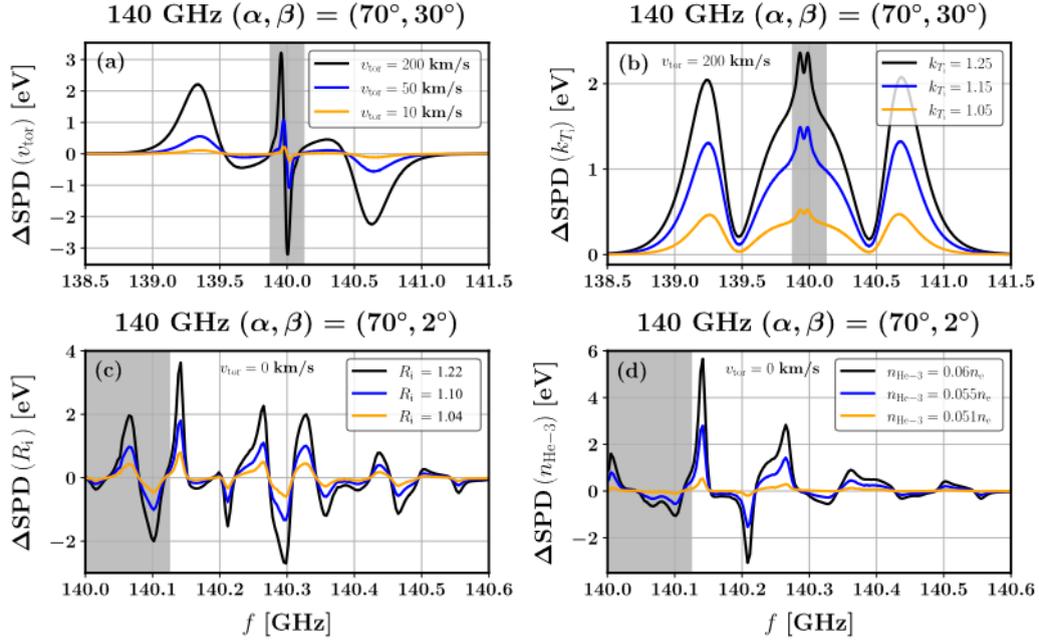

**Figure 9.** As fig. 6, but showing differences in the total CTS signal for varying plasma assumptions within the SV. (a) Signal difference between the case of a finite $v_{\text{tor}}$ and a plasma with no toroidal rotation. (b) Corresponding differences resulting from scaling up the ion temperature in the SV by a factor $k_{T_i}$ compared to the baseline value of $k_{T_i} = 1 \rightarrow T_{i,\text{SV}} \approx 19.8\,\text{keV}$. (c) Sensitivity to changes in the fuel-ion ratio compared to the baseline value $R_i = 1$ (section 2.2). (d) Impact of changes in $^3$He concentration compared to the baseline value $n_{\text{He}-3} = 0.05 n_e$ (section 2.2).

A qualitatively similar conclusion applies when varying the $^3$He concentration $n_{\text{He}-3}$ by 20% compared to the reference value of $n_{\text{He}-3} = 0.05 n_e$ (section 2.2) as shown in fig. 9(d), suggesting a resolvable signal difference for such a change in $n_{\text{He}-3}$.

## 5. Equatorial port geometry

So far, we have considered the upper and lower ports as launch points for the probe and receiver, respectively. However, such a constraint is neither a functional requirement for a CTS diagnostic in SPARC, nor a strict advantage. In this section, a geometry utilizing the equatorial port will be highlighted to demonstrate the flexibility of a potential CTS diagnostic in SPARC.

Based on the geometry from [2], we here confine probe and receiver to the same equatorial port, adopting launch points at $R = 2.45\,\text{m}$ and $z \pm 0.15\,\text{m}$ and a 140 GHz gyrotron frequency. Forgoing a systematic optimization of scattering geometries as in section 3.2, we instead manually identified a useful perpendicular geometry with $(\alpha, \beta) = (15°, 2°)$. Figure 10 shows the associated raytracing results, and table 5 summarizes the resulting scattering parameters.

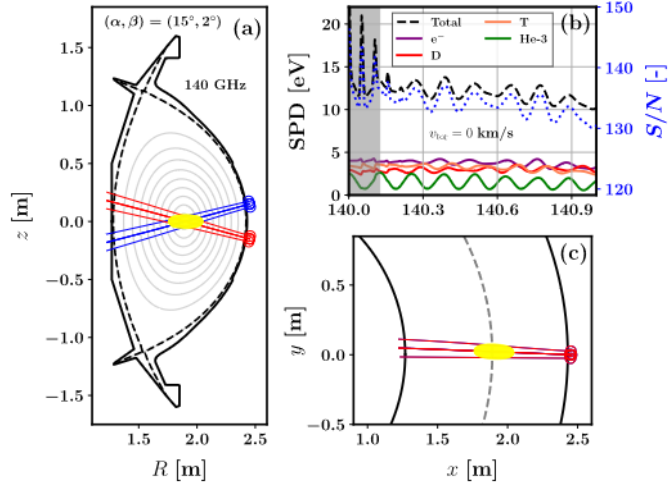

**Figure 10.** Perpendicular scattering geometry for core CTS measurements at 140 GHz utilizing an equatorial port, shown in a (a) poloidal and (c) toroidal cross section (see table 5). Panel (b) shows a resulting synthetic CTS spectrum and $S/N$, assuming $P^i = 200\,\text{kW}$, $v_{\text{tor}} = 0$, $\Delta f_b = 5\,\text{MHz}$, $\Delta t = 10\,\text{ms}$, and a noise level of $P^b = 2\,\text{eV}$. Grey region marks the frequency interval affected by the assumed notch filter.

**Table 5.** CTS parameters for a manually identified perpendicular 140 GHz geometry with probe and receiver beams launched from the same equatorial port.

| Equatorial port, perpendicular geometry | |
|---|---|
| $(\alpha, \beta)$ | (15°, 2°) |
| $\phi$ | 92.8° |
| $O_b$ [m$^{-1}$] | 25.2 |
| $\alpha_{\text{SP}}$ | 3.5 |
| $\rho_p$ | 0.05 |
| $\Delta R$ [m] | 0.23 |
| $\Delta z$ [m] | 0.06 |
| $\Delta \rho_p$ | 0.21 |

The identified equatorial-port geometry allows a centralized SV in the plasma. The volume of the SV is $\approx 463\,\text{cm}^3$, similar to that of the perpendicular 140 GHz geometry from table 2. However, due to the orientation of the SV, resulting from launch points with small vertical separation, the equatorial geometry has a smaller $\alpha_{\text{SP}}$ and a poorer spatial resolution in terms of $\Delta \rho_p$, but a larger overlap $O_b$. Nevertheless, both geometries share similar $\phi$, so the geometry of table 5 also implies sensitivity to variations in $R_i$ and $n_{He-3}$ at levels comparable to those in fig. 9. Finally, refraction remains limited for this geometry, as is apparent from fig. 10.

A synthetic spectrum for the equatorial geometry is shown in fig. 10(b). Even 1 GHz away from the probe frequency, the total SPD remains $> 10$ eV with $S/N > 100$. The oscillations in the $^3$He signal clearly impact the total signal, although the D and T signals remain dominant. While the peak total SPD is smaller than that of the 140 GHz spectrum shown in fig. 6(e), the signal in fig. 10(b) has a broader characteristic width.

Hence, locating probe and receiver in the same equatorial port represents a viable option for a CTS diagnostic designed to infer fuel-ion ratio or the $^3$He concentration in SPARC. A preliminary study suggests that a non-perpendicular geometry using these probe and receiver launch points is also feasible. Such a geometry would be sensitive to changes in $v_{tor}$ and $T_i$.

## 6. Discussion

Based on our results, we consider 140 GHz to be the optimal operating frequency for a SPARC CTS diagnostic among the three considered "off-the-shelf" gyrotron frequencies. A 140 GHz X-mode diagnostic would provide useful signal and $S/N$ levels for the inference of ion dynamics, while maintaining good spatial and temporal resolution, and limiting the sensitivity to refraction. The case of 175 GHz entails an ECE noise level up to a factor $10^3$ larger than at 105 or 140 GHz. On the other hand, refraction, which might hamper isotope CTS measurements, is significant at 105 GHz but limited at 140 and 175 GHz. Our recommended setup is thus reminiscent of that at the FTU tokamak, where a sub-harmonic 140 GHz X-mode diagnostic was operated in a high-field device [8, 44, 45].

With a single probe and receiver, a CTS diagnostic could measure either thermal- and fast-ion distribution functions or plasma composition. Measuring all quantities with a single CTS probe would require either steerable mirrors or two CTS receivers. In either of these cases, our results highlight the potential of a CTS diagnostic to complement the overall diagnostic suite at SPARC [46] by providing the following measurement capabilities:

★ Core-localized ($\rho_p \approx 0.05$) density measurements of at least D, T and $^3$He. This could complement edge measurements of neutral D, T, $^3$He, and $^4$He ash (and potentially other ion species) based on optical spectroscopy.

★ Core-localized measurements of main-ion $T_i$ and $v_{tor}$, complementing line-integrated impurity-ion measurements of these quantities using X-ray spectroscopy.

★ Core-localized measurements of either $T_i$ and the confined fusion $\alpha$ distribution function or of the fuel-ion ratio using a single CTS probe and receiver. With

steerable mirrors, all three quantities would be measurable in the plasma core. This would complement line-integrated measurements of $R_i$ and $T_i$ from neutron spectroscopic cameras [47] that also provide information on the neutron energy profile and the fusion-born $\alpha$ distribution.

Additionally, a CTS receiver could assist in detecting the occurrence of ELMs in SPARC through continuous monitoring of the ECE background at high temporal resolution.

This study has focused on probe and receiver launch points located in the upper and lower ports, respectively. As demonstrated in section 5, alternative geometries are also viable. However, in terms of integration into SPARC, we note that the vessel wall on the high-field side consists of vertical segments in the midplane, joined by upper/lower slanted segments that connect to the divertors, see, e.g., fig. 1. Due to their limited thickness, the vertical segments should not be exposed to large heat fluxes such as that of a largely unabsorbed gyrotron beam, even if operated at the relatively low power of 200 kW considered here. This risk is indeed eliminated by placing probe and receiver launch points in the upper and lower ports (fig. 4), whereas locating these in an equatorial port will result in the CTS probe beam terminating on a vertical segment (fig. 10). The latter setup would necessitate mitigating actions, such as operating the probe at low duty cycle and/or with a beam shape that ensures a large footprint and hence low power load on the wall.

While we have demonstrated the sensitivity of the considered CTS geometries to the plasma parameters discussed above, a more detailed study would be required to fully assess the accuracy with which these parameters can be measured by CTS. This would involve inversion of the synthetic spectra including noise as in [39, 43], which would also provide insight into suitable choices for $\Delta f_b$ and $\Delta t$. This study should furthermore assess the impact of microwave refraction and dispersion on the synthetic spectra [43], and could be expanded to include the predicted impact of edge plasma turbulence on the measurements.

A more detailed analysis would also be required to assess whether the assumed diagnostic beam shapes (including $w_0$ and $s_0$) could be optimized for improved spatial resolution and increased $O_b$ while satisfying integration constraints, e.g., in terms of size of diagnostic first-wall apertures.

Finally, our analysis suggests that the total diagnostic noise levels at 105 and 140 GHz are not dominated by ECE noise, in contrast to the case of existing CTS diagnostics. Rigorous quantification of any non-ECE noise, such as that from thermal plasma bremsstrahlung, may thus be required. Calculating the bremsstrahlung contribution at frequencies below the O-mode cutoff in SPARC ($\approx 180$ GHz) is non-trivial. While preliminary estimates suggest that this contribution would not alter our conclusions from section 3.3, this issue would need to be addressed in detail for a more mature diagnostic design.

# 7. Conclusions and outlook

In this paper, several conceptual designs for a CTS diagnostic in SPARC have been investigated.

Using the plasma profiles and magnetic equilibrium of the SPARC PRD scenario, sub-harmonic probing frequencies of 105, 140 and 175 GHz have been considered for probe and receiver in X-mode polarization. Gyrotrons operating at these frequencies are already in use at other devices and were chosen based on their relatively high technological readiness. A raytracing sweep of poloidal and toroidal injection angles was performed with the probe and receiver launch points located in an upper and lower port, respectively. This identified suitable scattering geometries, based on which we computed synthetic CTS spectra and signal-to-noise ratios.

While a diagnostic operating at 105 or 175 GHz would provide advantages in terms of low diagnostic background or insensitivity to refraction, respectively, a 140 GHz X-mode diagnostic is our recommended option. Such a setup can deliver core-localized measurements ($\rho_p \approx 0.05$) with good spatial resolution (radial extent $\Delta R \approx 8$ cm), robustness against refraction, low estimated background noise ($\lesssim 1$ eV), and high signal-to-noise ratios at reasonable integration times of $\Delta t \approx 20$ ms for an assumed gyrotron power of $P^i = 200$ kW.

Specific measurement capabilities include spatially resolved inference of fusion $\alpha$ densities (at an estimated accuracy of $\sim 10\%$, corresponding to that of ITER CTS) and the $\alpha$ distribution function, temperature and toroidal rotation of bulk D and T, fuel-ion ratio, and $^3$He concentration (relevant for the planned ion cyclotron minority heating scheme in SPARC). Core-localized measurements of these quantities are challenging to obtain by other means, particularly in the absence of neutral beam injection, so CTS could provide a strong complement to the already planned SPARC diagnostics such as X-ray, optical, and neutron spectroscopy.

However, simultaneous measurements of all the above quantities would require either steerable mirrors or two separate CTS receivers. We also emphasize that we have here only demonstrated good sensitivity to variations in the above parameters for a conceptual diagnostic design. A full assessment of the potential measurement accuracy would require a detailed analysis involving inversion of synthetic, noisy CTS spectra based on a more mature diagnostic design.

The proposed diagnostic configuration should, in principle, be feasible to integrate within the vessel/port layout and the diagnostic power budget. Relying on already available gyrotron designs, such an implementation may be possible at relatively low cost and with a short implementation time, in keeping with the short timeline for SPARC construction and commissioning.

## Acknowledgments

The information, data, or work presented herein builds on the SPARC primary reference discharge and X-point target discharge data provided by Commonwealth Fusion Systems [21].